# CMS Pixel Detector Upgrade

N. Parashar
*Purdue University Calumet, Hammond, IN 46323, USA*
*On behalf of CMS Collaboration*

The present Compact Muon Solenoid silicon pixel tracking system has been designed for a peak luminosity of $10^{34} cm^{-2}s^{-1}$ and total dose corresponding to two years of the Large Hadron Collider (LHC) operation. With the steady increase of the luminosity expected at the LHC, a new pixel detector with four barrel layers and three endcap disks is being designed. We will present the key points of the design: the new geometry, which minimizes the material budget and increases the tracking points, and the development of a fast digital readout architecture, which ensures readout efficiency even at high rate. The expected performances for tracking and vertexing of the new pixel detector are also addressed.

## 1. Introduction

The CMS detector at the LHC is located in Geneva, Switzerland. It is a general-purpose detector that is designed to operate at 14 TeV center of mass energy, delivered by proton-proton collisions at a peak luminosity of $10^{34} cm^{-2}s^{-1}$. Closest to the collision pipe are the tracking devices, which comprise the silicon strip detectors and the pixel detector [1]. These are placed inside a 4 Tesla solenoid magnet. At the heart of this is the pixel detector, which consists of two parts, namely the barrels in the central region and disks in the forward region. The current pixel detector was commissioned in 2008 and is running successfully in the data taking. However, it is envisaged that a continuous running for a period of 3 years at the design luminosity will deem the innermost layer inefficient for vertex reconstruction. Additionally, the upgrade of the LHC luminosity is also foreseen, which presents an experimental as we as technological challenge in terms of large particle fluxes and high radiation levels. This would require a replacement of the current pixel detector, designed for an increased luminosity of $2 \times 10^{34} cm^{-2}s^{-1}$, in the first phase of the upgrade, called Super LHC Phase 1. This note presents the CMS Pixel upgrade plans for the Phase 1 upgrade.

## 2. The current CMS pixel detector

The goal of the current CMS detector is to provide a robust tracking with precise vertex reconstruction within a strong magnetic field of 4 Tesla. This will be the key element to address the full range of physics accessible at energy of 14 TeV. At the heart of the tracking lies the pixel detector. The main task of this detector is vertex finding and flavor tagging. An efficient tagging of particles like b-quark, c-quark and tau, requires tracking as close as possible to the interaction vertex. The high spatial particle track density close to innermost tracking makes it necessary to be composed of pixel devices. These would provide a fine granularity in three-dimensional space information for vertexing. The discovery of Higgs Boson and Supersymmetric particles [2] relies on b-tagging. The study of B-physics, CP-violation and Top Quark physics require Pixel Detector. The Pixel Detector has an active surface area close to one square meter, instrumented with approximately 66 million channels. It consists of two components - barrels and disks. The disks comprise the Forward Pixel detector. The pixels are 150x100 μm in size with n-on-n silicon sensors of 260-300 μm, bump-bonded to PSI 46 Read Out Chips (ROC).

The pixel barrel consists of 3 layers with the innermost layer located at 4.3 cm and the outermost layer at 11.0 cm from the interaction point. All the 3 layers compose 48 million pixels, 11520 ROCs and 1120 readout links. The forward pixel detector comprises of two disks on either side of the interaction point, located at z=±34.5, ±46.5 cm, with 18M pixels, 4320 ROCs and 192 Readout links. Figure 1 shows the current layout of the Pixel detector, which was installed into the rest of CMS in Summer 2008.

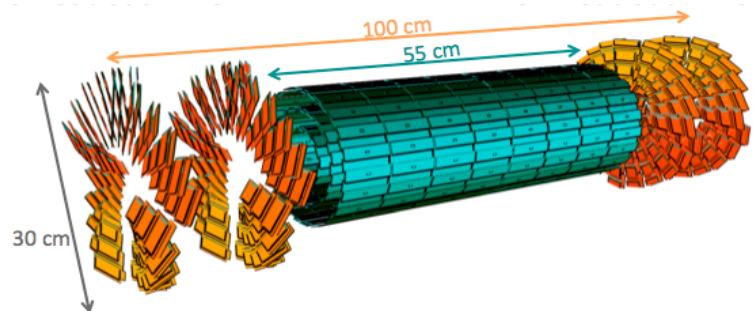

Figure 1: Layout of the current CMS pixel detector

## 3. Motivation for the Upgrade

The current pixel detector shall not be able to withstand the large occupancy resulting from the upgrade of the luminosity from $10^{34}$cm$^{-2}$s$^{-1}$ to $2\times10^{34}$cm$^{-2}$s$^{-1}$, in the Phase 1 upgrade of the SLHC. This increase in luminosity will primarily introduce readout inefficiencies with data losses due to limited buffer sizes. For e.g in layer 1 of barrel pixel alone, such looses are 16% with a 50 pile-up and 50% with a 100 pile-up. Therefore, one of the goals for an upgraded pixel detector would be to maintain a high hit detection efficiency, by making changes to the front-end electronics and data links. It will be equally crucial to maintain a good track seeding and pattern recognition performance, by assuring 4-hit coverage over the entire pseudo rapidity range, provide high track impact parameter resolution by reducing the passive material in the tracking volume, reducing the radius of the innermost layer and increasing radial acceptance. It will become very important to address the radiation damage issue for late Phase 1 run luminosity changes. While the current pixel detector can be replaced during the 2016-planned shutdown of the LHC, the option of an upgrade is a more viable solution for higher luminosity.

## 4. Layout for the proposed Upgrade

The proposed upgraded pixel detector consists of four barrel layers and three disk on either side of the interaction point. Figure 2 shows the layout of the new proposed design, with new geometry, new readout chip with enhanced features, a new cooling system based on two-phase $CO_2$, and a new powering system based on DC-DC converters [3].

The barrel pixel upgrade will consist of four cylindrical layers, with the innermost layer placed at 3.9 cm and the outermost layer at 16.0 cm, 1200 modules with 80 million pixels. The module used is only one type, a 8x2 ROC. The total material for the mechanical layout, including the support structure, cooling pipes and $CO_2$ is reduced from 400g to 100g. These changes yield a material budget reduction by a factor of 2. Figure 3 shows the material budget reduction in terms of radiation length. The Forward pixel upgrade will have 3 disks on each side of the barrel, with two concentric rings, inner and outer to facilitate easy replacement. These disks will utilize the same 8x2 module as the barrel, with 672 modules and 45 million channels in total.

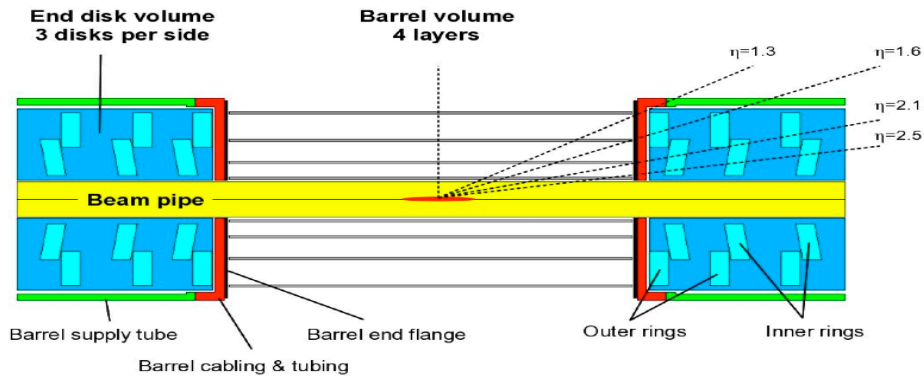

Figure 2: Layout of the proposed upgraded pixel detector with 4 barrel layers and 3 forward disks on each side.

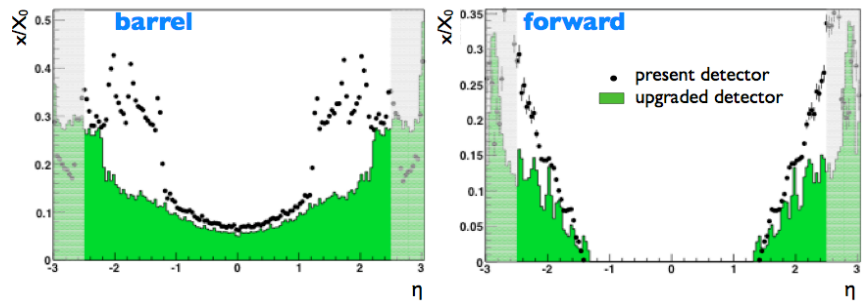

Figure 3: Material budget for the barrel and forward pixel detector. Radiation length of current barrel pixel (left) and forward pixel (right) detector shown in dots and proposed upgrade system shown as a histogram. The grey bands show the material distribution outside of the fiducial tracking volume.

## 5. Upgrade Simulation Studies

Simulation studies have been performed on the above-described plan for the CMS pixel upgrade option [3]. Simulation of the Phase I pixel upgrade show a significant improvement in tracking efficiency and reduced fake rates as demonstrated in Figure 4. Similarly the primary vertex resolution (Figure 5) and impact parameter resolution (Figure 6) improves significantly due to the increased sampling and reduced material.

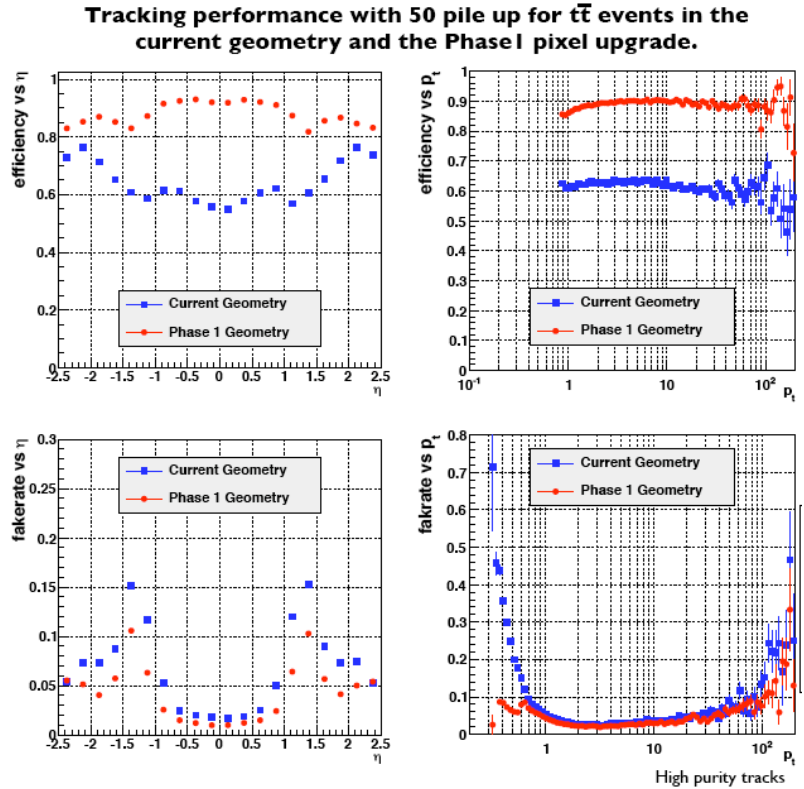

Figure 4: Comparison of the tracking efficiency as function of pseudorapidity, pT and fake track rate as a function of pseudorapidity and pT for the current (blue) and upgraded (red) detectors in events at $2 \times 10^{34}$ cm$^{-2}$s$^{-1}$ with 25 ns bunch spacing.

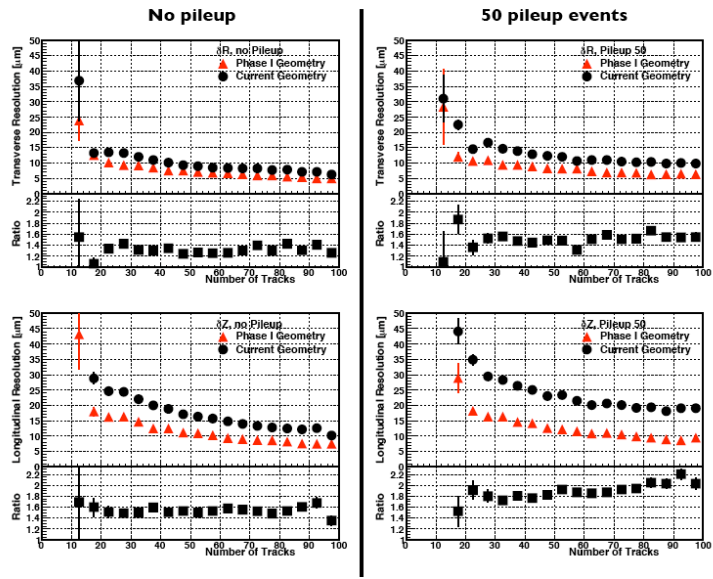

Figure 5: The transverse and longitudinal primary vertex resolutions for the present and upgraded versions of the CMS pixel detector as function of number of tracks in simulated top pair events.

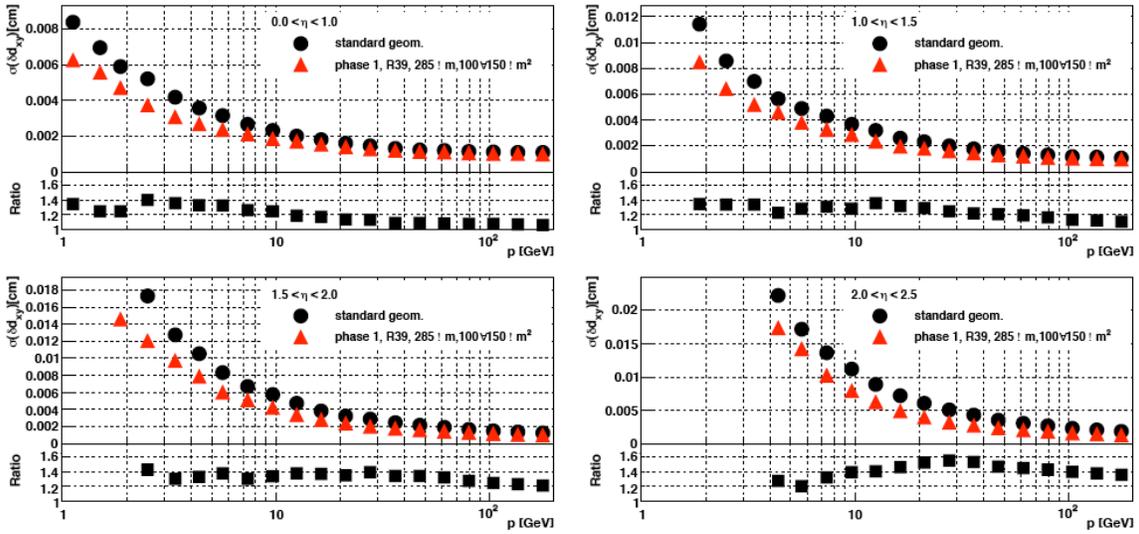

Figure 6: The longitudinal σ (δ$d_z$) impact parameter resolutions for the present and upgraded versions of the CMS pixel detector as function of track momentum in different pseudorapidity regions.

We expect the b-tagging performance to significantly improve in the Phase 1 upgrade (Figure 7), with the improved performance in track efficiency, track impact parameter, and primary vertex resolution.

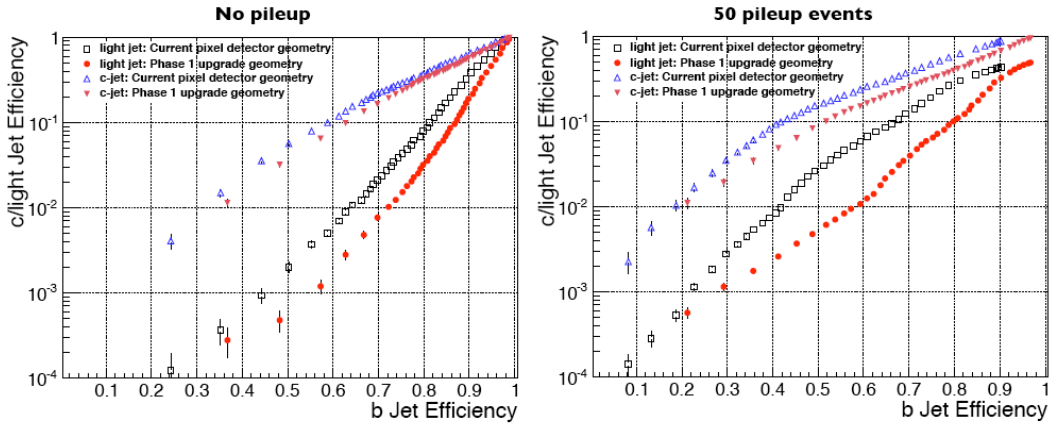

Figure 7: The b-tagging efficiency of the Combined Secondary Vertex Tagger is plotted versus the light quark (and gluon) efficiency for a sample of events in two different luminosity scenarios. The black points represent the performance of the current tracker and the red points represent the performance of the Phase 1 pixel upgrade. (a) The instantaneous luminosity is assumed to be low enough that there are no pile-up collisions (left). (b) The instantaneous luminosity $2 \times 10^{34} cm^{-2} s^{-1}$ with 25 ns bunch spacing (right).

## 6. Conclusions

The CMS pixel detector is installed, commissioned and operating as per the expectations. However, it will not survive the planned upgrade of the increase in luminosity in 2016. Therefore, a new upgraded pixel detector is a viable solution. The design of such a pixel detector is underway. The upgraded option will have new geometric layout with enhanced readout chip features, reduced material budget, a new cooling system and power system. Simulation studies related to tracking efficiency, primary vertex resolution, track impact parameter resolution and b-tagging performance presented in this note guide us that the upgraded option will be suitable to accommodate the Phase 1 upgrade of the SLHC.

## Acknowledgments

The author would like to thank National Science Foundation for their support.